\documentclass[smallextended]{svjour3}
\usepackage[dvips]{graphicx}

\journalname{Quantum Information Processing}

\begin{document}

\title{Any Quantum Network is Structurally Controllable by a Single Driving Signal}

\author{Michael Siomau}

\institute{M.~Siomau \at Physics Department, Jazan University,
P.O.~Box 114, 45142 Jazan, Kingdom of Saudi Arabia \\
\email{m.siomau@gmail.com}}

\date{Received: date / Accepted: date}

\maketitle

\begin{abstract}
Control theory concerns with the questions if and how it is possible
to drive the behavior of a complex dynamical system. A system is
said to be controllable if we can drive it from any initial state to
any desired state in finite time. For many complex networks, the
precise knowledge of system parameters lacks. But, it is possible to
make a conclusion about network controllability by inspecting its
structure. Classical theory of structural controllability is based
on the Lin's structural controllability theorem, which gives
necessary and sufficient conditions to conclude if a network is
structurally controllable. Due to this fundamental theorem we may
identify a minimum driver vertex set, whose control with independent
driving signals is sufficient to make the whole system controllable.
I show that Lin's theorem does not apply to quantum networks, if
local operations and classical communication between vertices are
allowed. Any quantum network can be modified to be structurally
controllable obeying a single driving vertex. \keywords{Quantum
Networks \and Controllability \and Linear Quantum Dynamics}
\PACS{03.67.Ac \and 89.75.Fb}
\end{abstract}

Modern science is more diverse than ever. Seemingly distant terms
such as quantum, network, learning, neural, complex and cryptography
are now combined into all-new scientific disciplines. While quantum
cryptography is a present technology \cite{Liao:17}, the study of
the structure and dynamics of complex quantum networks
\cite{Kimble:08} is at infancy. These networks are radically
different from their classical counterpart due to quantum
superposition and nonlocality, and unique features of quantum
dynamics and measurements \cite{Nielsen:00}. Quantum networks
exhibit non-classical clustering \cite{Pers:10} and synchronization
\cite{Manz:13}, and may undergo non-trivial phase transitions
\cite{Acin:07,Siomau:16}.

Whenever a complex system is concerned, be it classical or quantum,
we are likely to think if it's useful. And it surely is, if we can
predict its behavior and control it. The Lin's structural
controllability theorem \cite{Lin:74} is a bedrock of modern control
theory \cite{Liu:16}, which severely restricts our ability to gain
control over a complex system. I show that this restriction is no
longer valid for quantum networks, if we allow local operations and
classical communication (LOCC) between the network vertices.
Following general geometrical consideration, I show that a quantum
network with an arbitrary structure can exhibit structural
controllability with a single driving vertex, if modified with a
polynomial number of LOCC.

In classical control theory, if a complex system $(\bf{A};\bf{B})$
can be described with a set of linear differential equations
\begin{equation}
\label{linear system}
  \dot{\bf{x}}(t) \, = \, \bf{A} \, \bf{x}(t) + \bf{B} \, \bf{u}(t)
\end{equation}
at any time, it is called Linear Time-Invariant system
\cite{Liu:16}. Here, $\bf{x}(t)$ is the vector of system parameters;
$\bf{u}(t)$ is the input vector of independent driving signals; the
state matrix $\bf{A}$ describes which system components interact
with each other and the direction of the interaction; the input
matrix $\bf{B}$ identifies externally driven system parameters.

\begin{figure}
\begin{center}
\includegraphics[scale=0.23]{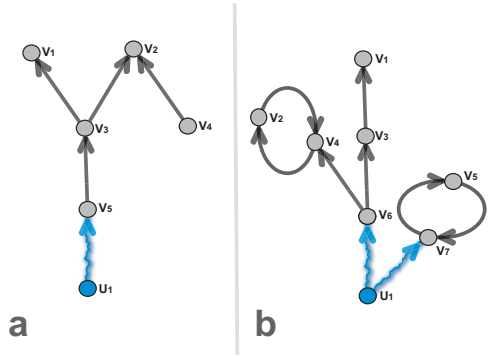}
\caption{(a) A digraph with a single root. Network vertices and
edges are shown in grey; driving vertices and edges - in blue.
Vertex $\{V_4\}$ is inaccessible from the driving vertex $\{U_1\}$.
Vertices $\{V_1, V_3, V_2\}$ form a dilation. (b) Elementary path
$\{ V_6, V_3, V_1 \}$ and two elementary cycles $\{ V_2, V_4\}$ and
$\{ V_5, V_7\}$ connected to the root $\{U_1\}$ form a cactus. This
network is structurally controllable by a single drive.}
 \label{fig 1}
\end{center}
\end{figure}

Such a system may be represented with a directed graph (digraph)
$G(\bf{A};\bf{B}) = (V_G;E)$, which structure doesn't change in
time. The vertex set $V_G = V \bigcup U $ includes both the state
vertices $V$ corresponding to the $N$ vertices of the network, and
the driving vertices $U$, corresponding to the $M$ input signals
that are called the  roots of the digraph $G(\bf{A};\bf{B})$. The
edge set $E = E_V \bigcup E_U$ consists of the edges among state
vertices $E_V$, corresponding to the connections of the network, and
the edges connecting driving vertices to state vertices $E_U$. In
this terms, Lin's theorem is given as: The system $(\bf{A};\bf{B})$
is {\it not} structurally controllable if and only if it has {\it
inaccessible nodes} or {\it dilations} \cite{Lin:74}.

A state vertex is inaccessible if there are no directed paths
reaching it from the input vertices. An inaccessible vertex can not
be influenced by driving signals, making the whole network
uncontrollable. The digraph $G(\bf{A};\bf{B})$ contains a dilation
if there is a subset of vertices $S \subset V$ such that the
neighborhood set of $S$ has fewer vertices than $S$ itself. Roughly
speaking, dilations are subgraphs in which a small subset of
vertices attempts to rule a larger subset of vertices (see
Fig.~\ref{fig 1}a).

This formulation is not practical, because doesn't tell us how many
driving signals we should have in a given network to make it
controllable. Alternatively, we may state that: An LTI system $(\bf{
A} ; \bf{B} )$ is structurally controllable if and only if
$G(\bf{A};\bf{B})$ is {\it spanned by cacti}.

A graph is {\it spanned} by a subgraph if the subgraph and the graph
have the same vertex set. For a digraph, a sequence of oriented
edges $\{v_1 \rightarrow v_2, ... , v_{k-1} \rightarrow v_k \}$,
where vertices $\{ v_1, v_2, ... , v_{k-1}, v_k \}$ are distinct, is
called an {\it elementary path} $C$. When $v_k$ coincides with
$v_1$, the sequence of edges is called an { \it elementary cycle}
$O$. For the digraph $G(\bf{A};\bf{B})$, let me define the following
subgraphs: (i) a stem is an elementary path originating from an
input vertex; (ii) a bud is an elementary cycle $C$ with an
additional edge $e$ that ends, but does not begin, in a vertex of
the cycle; (iii) a cactus is defined recursively: a stem $C$ is a
cactus. Let $C$, $O$, and $e$ be, respectively, a cactus, an
elementary cycle that is disjoint with C, and an arc that connects
$C$ to $O$ in $G(\bf{A};\bf{B})$. Then, $C \cup \{e\} \cup O$ is
also a cactus. {\it $G(\bf{A};\bf{B})$ is spanned by cacti if there
exists a set of disjoint cacti that covers all state vertices.}

\begin{figure}
\begin{center}
\includegraphics[scale=0.23]{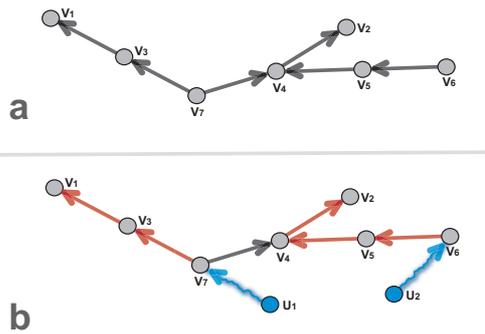}
\caption{(a) A graph to find a minimum driver set. (b) Maximum
matching is shown in red. Only two vertices $\{ V_6 \}$ and
$\{V_7\}$ are unmatched. The network requires two independent
drives/roots to be controllable.}
 \label{fig 2}
\end{center}
\end{figure}

A cactus is a minimal structure that contains neither inaccessible
nodes nor dilations (see Fig.~\ref{fig 1}b). A complex network is
unlikely to be spanned by a single cactus. But, it may be spanned by
a few. Since the control of a single cactus requires a single root,
to control a complex network we need as many driving signals as many
cacti span the network. In practice, we want to find the minimal
number of the driving signals to control a given network -- the
so-called minimum input problem \cite{Liu:16}. Hence, we need to
find the minimal number of cacti to span a given network. At first
glance, this combinatorial problem is a NP-hard, but in fact can be
resolved in polynomial time with {\it the maximum matching
algorithm} (see Fig.~\ref{fig 2}).

In a digraph, a matching is defined to be a {\it set of directed
edges} that do not share common start or end vertices. A vertex is
matched if it is the end vertex of a matching edge. Otherwise, it is
unmatched. Maximum matching is a matching of the largest size. This
definitions allow us to formulate the Minimum input theorem: To
fully control a system $G(\bf{A};\bf{B})$, the minimum number of
driver vertices is $N_D = {\rm max} \{ N - M, 1 \}$, where $M$ is
the size of the maximum matching in $G(\bf{A};\bf{B})$. In other
words, {\it the driver vertices correspond to the unmatched
vertices}. If all vertices are matched $M=N$ (as in case of an
elementary cycle), we need at least one input to control the
network, hence $N_D = 1$. We can choose any vertex as our driver
vertex in this case. Note, that in general the maximum matching is
not unique, i.e. the network may be controllable with different
minimal sets of driver vertices. But, all these minimal sets are of
the same size. The algorithm gives us the number and location of the
drivers to apply to the network, hence concludes the study of the
structural controllability.

In quantum networks, a vertex possesses a quantum system, such as
atom, quantum dot or molecule. If two vertices are connected with an
edge, they may communicate photons and classical information. An
edge is directed according to the ability of a vertex to send
photons to others. Classical information may be communicated between
connected vertices both ways irrespective of edge directions. This
setup fits most quantum communication protocols \cite{Nielsen:00}.

Because the quantum systems interact with each other and the
environment, that may be present in the vertices, the quantum
network dynamics is the subject of open quantum system dynamics
\cite{Gardiner:00}, which is in general non-linear. However, there
are few occasions, when the dynamics of an open quantum system may
be described with linear differential equations as Eq.~(\ref{linear
system}). For example, dissipative harmonic oscillator may be
described with linear Master and Langevin equations
\cite{Gardiner:00}. This type of quantum dynamics of the vertices we
must imply to ensure that the quantum network can be described with
Eq.~(\ref{linear system}). Whenever the behavior of a quantum system
cannot be correctly described with a system of linear differential
equations (\ref{linear system}), further considerations are invalid.

The difference between classical and quantum networks is that the
latter are non-local. This leads to the fact that distant vertices
may become entangled and display correlated dynamics, even so they
are not connected with an edge. The entanglement may appear
spontaneously as a result of the non-linear dynamics \cite{Manz:13}
or may be created by LOCC \cite{Acin:07}. The entanglement between
distant vertices may be interpreted as a new edge of the network
\cite{Acin:07,Siomau:16}. The direction of this entanglement edge is
defined by the direction of the classical communication. The LOCC
may be used to loop all inaccessible vertices and dilations into
elementary cycles, modifying the network so, that it is spanned by a
single cactus, i.e. all its vertices are matched.

To be specific, let us suppose that we have a connected quantum
network represented by a digraph and of the same configuration as in
Fig.~\ref{fig 2}a. We may investigate its classical structural
controllability by executing the maximum matching algorithm to
identify the minimal set of driving vertices. The network is spanned
by two cacti, due to a dilation $\{V_3, V_7, V_4\}$. Let us choose
one of the two elementary pathes, let say $\{V_6, V_5, V_4, V_2\}$,
and loop it into an elementary cycle by LOCC.

The procedure can be exemplified as follows. Suppose we have a
quantum network consisting of three vertices $\{a, b, c\}$, so that
$a$ is connected to $b$ and $b$ is connected to $c$. The quantum
system at the vertex $a$ may be entangled with a quantum field mode,
which is communicated to the vertex $b$. Subjected with the
classical communication vertex $a$ is entangled with $b$, i.e. we
created an entanglement edge. Through this edge, the quantum system
at $b$ may influence the system at $a$ by performing local
manipulation at $b$ and communicating information classically to $a$
\cite{Nielsen:00}. In this case, the direction of the classical
communication sets the direction of the entanglement edge. If $a$ is
entangled to $b$ and $b$ is entangled to $c$, we may apply
entanglement swapping at $b$. This creates a non-local entanglement
edge $a-c$. The edge may be directed both ways depending on the
classical information flow.

Similarly, we may get rid of inaccessible vertices, in case of their
presence. If we have a subgraph of inaccessible vertices in our
connected graph, then there is at least one border vertex of the
subgraph that has an edge directed from it to an accessible vertex.
The inaccessible vertex may become entangled with the accessible
vertex by LOCC. Hence, this two vertices are looped into an
elementary cycle consisting of just two vertices. Iteratively, all
the subgraph of inaccessible vertices may become accessible. As the
result, the modified network is spanned by a single cactus, thus
controllable by a single root. In general, the whole procedure can
be executed for arbitrary two distant vertices in a finite connected
network of size $N$ with at most $N^3$ LOCC \cite{Siomau:AIP}.

\begin{figure}
\begin{center}
\includegraphics[scale=0.23]{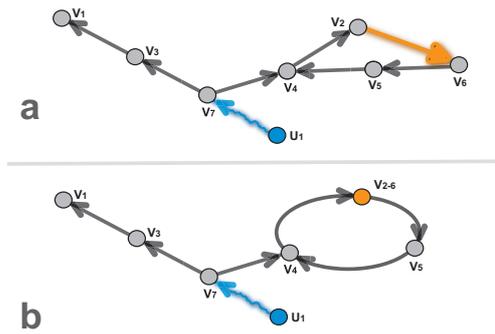}
\caption{(a) By LOCC we may loop the elementary path $\{V_6, V_5,
V_4, V_2\}$ into an elementary circle increasing maximal matching of
the quantum digraph. (b) The modification of the quantum digraph
with LOCC may be viewed as creation of a supervertex $\{ V_{2-6}
\}$.}
 \label{fig 3}
\end{center}
\end{figure}

I showed that the restrictions of the Lin's theorem, that are
fundamental for classical networks, doesn't apply to quantum
networks. This is another radical difference between classical and
quantum instances. The result is independent of the quantum network
structure and is purely based on our ability to create nonlocal
correlations with LOCC.

To make a conclusion about structural controllability of a network,
we must be sure that the structure of the network doesn't change in
time. This may be guaranteed if the network is described with a
system of linear equations. But, do LOCC change linearity of the
quantum dynamical equations? Under some reasonable assumptions, the
initial entanglement between two systems leaves the dynamical
equations linear adding an inhomogeneous term \cite{Stelm:01}. But,
this is an exception rather than the rule. Can we still conclude
about the structural controllability? Yes, if we consider the
entangled vertices as single super-vertex (see Fig.~\ref{fig 3}).
Then, we must ensure that the super-vertex is coupled linearly to
the neighbors, i.e. the output field operators of the super-vertex
and the input field operators of the neighboring vertices are
linearly dependant \cite{note}. In this case the network remains
linear, in spite of the non-liner internal dynamics of the
super-vertex.

Structural controllability analysis is the very first step to
evaluate network controllability. Although a network may be
structurally controllable, for some combination of the dynamical
parameters, it may not be controllable \cite{Liu:16}. Hence, we need
to establish the so-called strong structural controllability, that
is the network is controllable for any combination of the dynamical
parameters. This analysis requires full information about the
network parameters and analytical tools to process this information.
The Kalman's criterion of controllability \cite{Liu:16} is as
fundamental as the Lin's theorem for structural controllability, and
may also fail as the latter, if applied to quantum networks. This
urges a revision of our knowledge and strategies to establish
control over complex quantum networks.

\end{document}